\documentclass[a4paper,showpacs,twocolumn,amsmath,amsart,amssymb,floatfix,prb,preprintnumbers,footinbib,superscriptaddress]{revtex4-1}
\usepackage{graphicx}
\usepackage{amsmath,amsfonts}
\usepackage{color}

\newcommand{\G}{\Gamma}

\newcommand{\e}{\varepsilon}

\newcommand{\si}{\sigma}
\newcommand{\W}{\boldsymbol{\mathcal{W}}}
\renewcommand{\P}{\boldsymbol{p}}

\begin{document}

\title{Thermoelectric performance of a driven double quantum dot}

\author{Stefan Juergens} 
\affiliation{Institut f\"ur Theorie der Statistischen Physik, RWTH Aachen University, D-52056 Aachen, Germany}
\affiliation{  JARA-Fundamentals of Future Information Technology}
\author{Federica Haupt} 
\affiliation{Institut f\"ur Theorie der Statistischen Physik, RWTH Aachen University, D-52056 Aachen, Germany}
\affiliation{  JARA-Fundamentals of Future Information Technology}
 \author{Michael Moskalets} 
\affiliation{Department of Metal and Semiconductor Physics, NTU ``Kharkiv Polytechnic Institute", 61002 Kharkiv, Ukraine}
\author{Janine Splettstoesser} 
\affiliation{Institut f\"ur Theorie der Statistischen Physik, RWTH Aachen University, D-52056 Aachen, Germany}
\affiliation{  JARA-Fundamentals of Future Information Technology}
\date{\today}
\begin{abstract}
In this paper we investigate the thermoelectric performance of a double-dot device driven by time-dependently modulated gate voltages. We show that if the modulation frequency $\Omega$ is sufficiently small, not only quantized charge pumping can be realized, but also the heat current flowing in the leads is quantized and exhibits plateaux in units of $\frac{\Omega}{2\pi}\, k_{\rm B}T \ln2$.
The factor ln2 stems from the degeneracy of the double-dot states involved into transport. This opens the possibility of using the pumping cycle to transfer heat against a temperature gradient or to extract work from a hot reservoir with Carnot efficiency. However, the performance of a realistic device is limited by dissipative effects due to leakage currents and finite-frequency operation, which we take into account rigorously by means of a generalized master equation approach in the regime where the double dot is weakly coupled to the leads. We show that despite these effects, the efficiency of a double-dot charge pump performing work against a dc-source can reach of up to 70\% of the  ideal value.
\end{abstract}


\maketitle

\section{Introduction}

\begin{figure}[h]
\includegraphics[width=3.in]{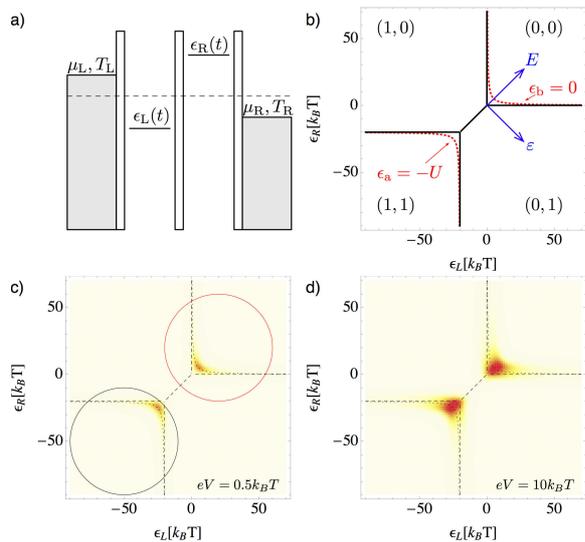}
\caption{(Color online)  (a) Sketch of the potential landscape of a double-dot setup. 
(b) Stability diagram of the double dot for the case $V=0$: black full lines indicate the borders of the stability regions for negligible inter-dot coupling, while the red dashed lines show the resonance positions of bonding and antibonding state. 
The system of coordinates formed by the detuning $\e=\epsilon_{\rm L}-\epsilon_{\rm R}$ and the mean energy $E=(\epsilon_{\rm L}+\epsilon_{\rm R})/2$ is also shown.
(c) Sketch of possible pumping trajectories in parameter space. The color-scale plot represents the dc charge-current  through the double dot in the linear response regime ($eV=0.5k_{\rm B}T$); (d)  Dc charge-current at finite bias ($eV=10k_{\rm B}T$). Other parameters: $U=20 k_{\rm B}T$, $t_{\rm c}=10 k_{\rm B}T$, $\Gamma_{\rm L}=\Gamma_{\rm R}=\Gamma/2$, $\hbar\Gamma=k_{\rm B}T/4$ and $T_{\rm L}=T_{\rm R}=T$. 
\label{fig_setup}}
\end{figure}

The tunability of nanoscale systems such as quantum dots or single-electron boxes allows to exploit their functionalities in the realization of nanoscale electric and thermoelectric devices.  
While the use
of quantum dots and metallic islands as single-electron transistors,\cite{Kastner92}  ultra-sensitive detectors,\cite{Clerk10}  or thermometers\cite{Giazotto06} has already been considered for a long time, it is only more recently that their applications as thermoelectric engines,\cite{Esposito09,*Esposito10,Sothmann12b,Whitney13} refrigerators,\cite{Edwards93,Humphrey01,Arrachea07,Rey07,Prance09,Venturelli12} and heat rectifiers \cite{Sanchez11,Ruokola11,Sothmann12,Sanchez12,Jordan13,DSanchez13} -- to name some examples --   started to attract considerable attention.  
Since Coulomb interaction and quantum interference effects play an important role in these devices, their thermoelectric properties differ strongly from those of classical macroscopic systems,\cite{Sivan86,Beenakker92,*Molenkamp94,Kubala06,*Kubala08,Kirchner12,Costi10,Kennes13} and can even result in a significant increase of the thermoelectric efficiency.\cite{Karlstrom11,Andergassen11,Trocha12,Meair13} Moreover, thanks to their stability and to the possibility of  monitoring their state with integrated charge or temperature detectors, quantum dots and single-electron boxes represent an ideal playground to experimentally test the predictions of so-called fluctuation theorems,~\cite{Averin11a,*Saira12,Kung12,Solinas13}  or to investigate possible implementations of Maxwell's demons in solid state environment.~\cite{Averin11,Strasberg13}

Our particular interest concerns the operation of a quantum dot system driven by time-dependent fields. Since the original experiment of Pothier {\em et al.}~\cite{Pothier92} it is well known that by changing slowly the gate voltages applied to a double dot (or a double metallic island as used in Ref.~\onlinecite{Pothier92}), electrons can be transferred from one lead to the other in a controlled way, generating a quantized dc-current $\bar{I}=\pm e\Omega/2\pi$, where $\Omega$ is the frequency of the ac-signal applied to the gates. This result has a clear interest for metrology,\cite{Pekola_rev} and stimulated an intense experimental\cite{Keller96,Chorley12dd,Roche13,Connolly12} and theoretical activity.\cite{Dalsgaard92,Kashcheyevs08}
Importantly, single-electron pumping can be achieved both when the two leads are in the same equilibrium state and when a finite bias voltage is applied to the device,  as well as it is possible to pump electrons {\em against} the preference direction set by the bias. 

Beyond the goal of realizing quantized charge pumping, there has been interest in time-dependently driven double-dot systems 
 thanks to their complex internal structure, which results in highly sensitive devices. To name some examples, time-dependently driven hybrid double-dot devices with ferromagnetic leads have been shown to allow for pure spin currents,\cite{Cota05,*Cota05E,Khomitsky09,Riwar10} the orbital degree of freedom of a double dot allows to study spin-orbit coupling effects in pumping,\cite{Romeo09,Rojek13} a double-dot pump with superconducting leads has been proposed as a detector for crossed Andreev reflection~\cite{Hiltscher11}, and Landau-Zener transitions in transport have been also studied.\cite{Speer00,Renzoni01,Roche13}

Motivated by the possibility of pumping charge against a bias, we study the performance of a driven double dot as a nanoscale ``battery charger'' transferring electrons from a lower to a higher chemical potential, as well as we investigate its efficiency as  a heat pump (or heat engine) operating between two reservoirs at different temperatures.  To this end, we consider the charge and heat currents flowing in the leads in response to the time-dependent driving. 
Heat currents in electronic  quantum pumps have been so far mostly investigated in the limit  where the electronic interactions can be neglected.\cite{Moskalets02,Wei04,Rey07,Arrachea07,Moskalets09,Battista12} The operation of the double-dot pump is  however based on charging effects,\cite{Pothier92}  and therefore we adopt  a generalized master equation approach,\cite{Konig96a,*Konig96b} which allows to take into account arbitrarily strong Coulomb interaction between electrons in the dots, as well as non-equilibrium conditions induced by dc voltage and temperature gradients, and the time-dependent driving.\cite{Splett06,Cavaliere09}  
We solve the dynamics of the system in terms of an adiabatic expansion for the reduced density matrix in the limit of weak coupling to the leads and slow driving. Contributions to the charge and the heat currents are taken into account up to the second non-adiabatic correction, which is particularly important in order to account properly for heating effects due to the ac-driving.  

We show that  in certain regimes not only the pumped charge current is quantized, but also the heat current shows well defined plateaux which are directly related to the electronic  temperature in the leads and the degeneracy of the double dot states involved into transport. Moreover, we show that in 
 these regimes the pumping cycle can be considered as a close analog of the Carnot cycle and that the double-dot pump can in principle be employed to transfer heat against a temperature gradient, to extract work from a hot reservoir, or to move charges from a lower to a higher chemical potential with maximal efficiency, if the driving is infinitely slow. The performance of a realistic pump is however limited by leakage currents and heat production due to finite-frequency operation. We investigate these limitations  in detail and find that, at least for the case of the charge pump working against a dc-bias, efficiencies up to 70\% of the maximal value can be obtained.

The manuscript is structured as follows. We introduce the model for the double dot and the generalized master equation approach used for the calculation of the charge and heat currents in Section~\ref{sec_model}. In Section~\ref{sec_ideal} we discuss first the case of pure adiabatic pumping of charge and heat, and then  present the operation of three ideal double dot based engines. Finally, the  limitations to the perfect operation of these engines due to leakage currents and finite-frequency driving are discussed in Section~\ref{sec_limitations}.

\section{Model and Technique}\label{sec_model}

\subsection{Double dot in the molecular regime}

We consider a double-dot device formed by two single-level spin-degenerate quantum dots connected in series and coupled to external leads, see Fig.~\ref{fig_setup}a. 
This system is described by the Hamiltonian
\begin{subequations}\label{eq:H}
\begin{eqnarray}
\hat{H} & = & \hat{H}_\mathrm{dd}+\sum_{\alpha=\rm L,R}\hat{H}_\alpha+\hat{H}_\mathrm{tun}.
\end{eqnarray}
Here, $H_\mathrm{dd}$ is the Hamiltonian of the isolated double dot:  
\begin{align}
\hat{H}_\mathrm{dd} =  &\sum_{\alpha=\mathrm{L},\mathrm{R}} \epsilon_{\alpha}\hat{n}_{\alpha}+U \hat{n}_\mathrm{L}\hat{n}_\mathrm{R}
+\frac{U'}2\!\!\sum_{\alpha=\mathrm{L},\mathrm{R}}\!\!\hat{n}_{\alpha}(\hat{n}_{\alpha}-1) \nonumber\\& 
 -\frac{t_{\rm c}}{2}\sum_{\sigma=\uparrow,\downarrow}\left(\hat{d}_{\mathrm{L}\sigma}^\dagger\hat{d}_{\mathrm{R}\sigma}^{}+\mathrm{h.c.}\right).
\end{align}
The dot operators $\hat{d}_{\alpha\sigma}^\dagger (\hat{d}_{\alpha\sigma}^{})$ create (annihilate) an electron with spin $\sigma=\uparrow,\downarrow$  and energy $\epsilon_\alpha$ in the dot $\alpha=\mathrm{L},\mathrm{R}$. The corresponding number operator of electrons in each dot is given by $\hat{n}_\alpha=\hat{n}_{\alpha\uparrow}+\hat{n}_{\alpha\downarrow}$ with $\hat{n}_{\alpha\sigma}=\hat{d}_{\alpha\sigma}^\dagger\hat{d}_{\alpha\sigma}^{}$; the total occupation-number operator of the double dot is $\hat{n}=\hat{n}_\mathrm{L}+\hat{n}_\mathrm{R}$. The inter- and intra-dot Coulomb interaction is denoted by $U$ and $U'$, respectively.  In the following, we will assume the onsite interaction $U'$ to be the largest energy scale in the system ($U' \to \infty$), so that each dot can be at most singly occupied.  Hopping from one dot to the other occurs with the inter-dot coupling amplitude $-t_{\rm c}/2$, where 
$t_{\rm c}$ is taken to be real and positive. 
The single-particle energies $\epsilon_\alpha=\epsilon_\alpha(t)$ of the dots 
can be tuned by external gates locally applied to the two dots and are in general time-dependent.

The leads are described as non-interacting Fermi liquids with the Hamiltonian
\begin{equation}
\hat{H}_\alpha=\sum_{k,\sigma=\uparrow,\downarrow}\epsilon_{\alpha k \sigma}\hat{c}_{\alpha k\sigma}^{\dagger}\hat{c}_{\alpha k\sigma}^{},
\end{equation}
where $\hat{c}_{\alpha k \sigma}^{\dagger}(\hat{c}_{\alpha k\sigma}^{})$ are the creation (annihilation) operators for an electron with momentum $k$ and spin $\sigma$ in lead $\alpha$. 
Finally, the coupling between the double dot and the leads is given by 
\begin{equation}
\hat{H}_\mathrm{tun}=\sum_{\alpha,k,\sigma}\left( t_\alpha\hat{c}_{\alpha k\sigma}^\dagger\hat{d}^{}_{\alpha\sigma}+\mathrm{h.c.}\right).
\end{equation}
\end{subequations}
The coupling is quantitatively  characterized by the energy-independent tunnel-coupling strength $\Gamma_\alpha=2\pi\nu_\alpha|t_\alpha|^2/\hbar$. Here $t_{\alpha}$ is  the tunneling amplitude between dot $\alpha$ and its neighboring lead, which we assume to be spin and momentum independent, and $\nu_\alpha$ is the density of states of lead $\alpha$ in the wide band limit.

The imposed boundary conditions are such that each lead is in local equilibrium with  the electrochemical potential $\mu_\alpha$ and the temperature $T_\alpha$. Here we take the mean electrochemical potential $\mu=\frac{\mu_L+\mu_R}{2}$ as the zero energy level, i.e. $\mu=0$. All single-particle energies in Eq.\eqref{eq:H} are expressed with respect to it. The difference between the electrochemical potentials of the two leads is fixed by the applied bias voltage $V$.  In the rest of the paper we will assume $\mu_{\rm L}=-\mu_{\rm R}=eV/2>0$, with $e>0$ the absolute value of the  electron charge. 

In the following we will focus on the case of strong inter-dot coupling, namely when the inter-dot dynamics is much faster than the dot-lead hopping ($t_c \gg \hbar \Gamma$ with $\Gamma=\Gamma_{\rm L}+\Gamma_{\rm R}$).
In this limit, it is useful to diagonalize the single-particle sector of $\hat{H}_{\rm dd}$ introducing its ``molecular" eigenstates, namely the bonding and antibonding states  $|{\rm b}\sigma\rangle=d^\dag_{\rm b\sigma}|0\rangle$ and $|{\rm a}\sigma\rangle=d^\dag_{\rm a\sigma}|0\rangle$, which go along with the creation operators
\begin{equation}\nonumber
\hat{d}^\dagger_{\mathrm{{\rm b}/{\rm a}}\ \sigma} \!\!= \! \frac{1}{\sqrt{2}}\!\!\left[\!\sqrt{1\!\pm\! \frac{\e}{\sqrt{\e^2+t_{\rm c}^2}}}\ \hat{d}^\dagger_{\mathrm{R}\sigma} \! \pm \!\sqrt{1\!\mp\!\frac{\e}{\sqrt{\e^2+t_{\rm c}^2}}}\ \hat{d}^\dagger_{\mathrm{L}\sigma}\!\right].
\end{equation}
These states have the energies
\begin{eqnarray}
\epsilon_\mathrm{{\rm b}/{\rm a}} & = & E\mp\frac{1}{2}\sqrt{\e^2+t_{\rm c}^2},
\end{eqnarray}
where the upper (lower) sign corresponds to the bonding (antibonding) state and $E=(\epsilon_\mathrm{L}+\epsilon_\mathrm{R})/2$ and $\e=\epsilon_\mathrm{L}-\epsilon_\mathrm{R}$ are the mean energy and the detuning between the energies of the two dots, respectively (see Fig.~\ref{fig_setup}b).  
In this new basis, the tunnel coupling between the double dot and the two leads is expressed by effective rates for tunneling via  the hybrid states $|\rm b\sigma \rangle $ and $|\rm a \sigma\rangle $:\cite{Riwar10}
\begin{subequations}\label{eq:effective_coupling}
\begin{align}
\Gamma_{{\rm L},{{\rm b}/{\rm a}}}&=\frac{\Gamma_{\rm L}}2\left(1\mp \frac{\e}{\sqrt{\e^2+t_{\rm c}^2}}\right),\\
\Gamma_{{\rm R},{{\rm b}/{\rm a}}}&=\frac{\Gamma_{\rm R}}2\left(1\pm \frac{\e}{\sqrt{\e^2+t_{\rm c}^2}}\right).
\end{align}
\end{subequations}
We also define the total broadening of the bonding and antibonding states  as  $\Gamma_{{\rm b}/{\rm a}}=\sum_{\alpha=\rm L,R}\Gamma_{\alpha,{{\rm b}/{\rm a}}}$.  
Because of the dependence on the level detuning $\e$,  these effective rates become a function of time whenever an out-of-phase modulation is applied to the gates of the two dots. 

The stability diagram of the double dot is shown in Fig.~\ref{fig_setup}b, and identifies regions of different equilibrium occupation numbers for the two dots, as a function of the energies $\epsilon_{\rm L}$ and $\epsilon_{\rm R}$. 
The points where three charge states are degenerate are named {\em triple points}. For strong inter-dot coupling $t_{\rm c}\gg\hbar\Gamma$, the edges of the stability regions are defined by the conditions $\epsilon_{\rm b}=0$ and $\epsilon_{\rm a}+U=0$ 
and an anti-crossing behavior is shown at the triple points. 
 This anti-crossing can be clearly distinguished in the dc-current in the linear response regime (see Fig.~\ref{fig_setup}c), and is a hallmark of a tunnel-coupled double-dot system. For finite bias the triple points get broadened (see Fig.~\ref{fig_setup}d), as there are now entire regions in parameter space that correspond to no stable charge configuration.

\subsection{Generalized master equation}\label{sec:kineticeq}

We want to describe the dynamics of the double dot in the presence of a time-dependent driving applied to its gates in the limit of weak coupling to the leads $\hbar\Gamma\ll k_\mathrm{B}T$. 
The state of the double dot at a given time $t$ is described in general by its reduced density matrix $\hat{\rho}_{\rm dd}(t)$. However, if the inter-dot coupling is much stronger than the one to the leads ($t_{\rm c}\gg\hbar\Gamma$), the bonding and anti-bonding states are non-degenerate even for zero detuning and 
the dynamics of the diagonal and off-diagonal elements of  $\hat{\rho}_{\rm dd}(t)$ decouple in  lowest order in $\Gamma$.~\cite{Leijnse08} In this case, we can restrict ourselves to study only the diagonal elements of $\hat{\rho}_{\rm dd}(t)$, namely the occupation probabilities.~\cite{Riwar10} 
Their evolution is governed by the generalized master equation\cite{Konig96a,*Konig96b}
\begin{equation}
\frac{d}{d t}\P(t)=\int^t_{-\infty}dt'{\bf W}(t,t')\P(t'), \label{eq:mastereq}
\end{equation}
with  $\P=\left(p_0,p_{\mathrm{a}\uparrow},p_{\mathrm{a}\downarrow},p_{\mathrm{b}\uparrow},p_{\mathrm{b}\downarrow},p_{\uparrow\uparrow},  p_{\uparrow\downarrow},  p_{\downarrow\uparrow},  p_{\downarrow\downarrow}\right)^{T}$, where we omitted the time arguments for simplicity. These are the probabilities that the double dot is empty, $p_0$, that it is singly occupied with an electron with spin $\uparrow$ or $\downarrow$ in the bonding  or in the antibonding state $p_{\mathrm{b}\sigma}$ and $p_{\mathrm{a}\sigma}$, or finally that the double dot is doubly occupied, $p_{\sigma\sigma'}$, where the two subscripts indicate the spin of the electrons occupying the left and right dot, respectively. As mentioned before, double occupation of a single dot is energetically forbidden due to strong onsite repulsion. The kernel ${\bf W}(t,t')$ is a transition matrix that incorporates the tunnel coupling to the leads. It has a functional dependence on the time-dependent parameters $\epsilon_\alpha(\tau)$, with $\tau\in [t,t']$.   

We focus here on the case of slow driving, where the life-time of the electrons in the system is much shorter than  the period of the driving. In this regime, Eq.\eqref{eq:mastereq} can be solved perturbatively by performing an 
adiabatic expansion for the occupation probabilities of the system $\P(t)\to\sum_{k\ge0}\P^{(k)}_{t}$. Here, $\P^{(0)}_{t}$ is the solution of the problem with all parameter values frozen at time $t$. It represents the steady state the system would relax into if it could instantaneously follow the modulation of the time-dependent parameters. We will therefore refer to it as the {\em instantaneous} solution. Corrections due to retardation effects are encoded in $\P^{(k>0)}_t$, and are governed by a competition between the time scales of the driving and the response times contained in $\bf W$. They are the solutions of a hyerarchy of equations\cite{Splett06} and, in the considered case of periodic driving, they 
are associated with different powers of the driving frequency, i.e. $\P^{(k)}_{t}\propto\Omega^{k}$.\footnote{Note that, the order-$k$ in our expansion is not related with the number of absorbed energy quanta within a Floquet approach.}
Alongside this adiabatic expansion, we perform a perturbative expansion in the strength of the coupling to the leads $\Gamma$. Details of such a double expansion can be found in Refs.~\onlinecite{Splett06, Cavaliere09}. In the limit of weak coupling to the leads ($\hbar \Gamma\ll k_B T$), retaining only terms to lowest order in $\Gamma$, this results in the set of equations\cite{Splett06,Cavaliere09}
\begin{subequations}\label{eq:hierarchy}
\begin{align}
&{ \W}_t\P^{(0)}_t=0,\\
&{ \W}_t\P^{(k)}_t=\frac{d}{dt}\P^{(k-1)}_t.\label{eq:masterPk}
\end{align}
\end{subequations}
Here, the matrix ${\W}_t$ is the zero-frequency Laplace transform of the kernel in time-space  with all time-dependent parameters frozen at
time $t$, ${{\W}_{t}=\int_{-\infty}^{t}d(t'-t){\bf W}(t-t')}$, evaluated to first order in $\Gamma$. It contains the transition rates between different double-dot states as given by Fermi's golden rule.    It satisfies the sum rule $\sum_{i}[\W_{t}]_{ij}=0$, which ensures the conservation of probability, and can be written as a sum of independent contributions from the two reservoirs ${\W}_t=\sum_{\alpha=L,R}\W_{\alpha,t}$.
Together with the normalization conditions ${\sum_{i}p_{i,t}^{(0)}=1}$ and ${\sum_{i}p_{i,t}^{(k>0)}=0}$, 
Eqs.\eqref{eq:hierarchy}  allow to evaluate iteratively all the non-adiabatic corrections $\P^{(k)}_t$.\cite{Splett06} Since the kernel ${\W}_t$ depends parametrically on time (as emphasized by the subscript $t$), also the solutions of Eq.\eqref{eq:hierarchy},  $\P^{(k)}_t$, acquire a parametric dependence on $t$.

We stress that the validity of Eq.\eqref{eq:hierarchy} is restricted to the regime of weak coupling to the leads $\hbar \Gamma\ll k_{\rm B}T$ and {\em slow driving}. More precisely the latter condition requires that the frequency and the amplitude of the modulation satisfy  the relation $\delta_{X}  \Omega\ll \Gamma k_\mathrm{B} T$, where $\delta_{X}$ represents the amplitude of any of the modulation parameters. In this limit, the expansion in powers of $\Omega$ for the occupation probability can be truncated to $k=2$, as it will be discussed throughout this paper.

\subsection{Charge and heat currents}
We are interested in the charge and heat currents flowing in the leads in the presence of an external time-dependent driving. For definiteness, we take the sign convention that in each lead the particle (heat) current is positive when flowing towards the double dot. With this convention, the {\em charge} current in lead $\alpha$ can be written as
\begin{equation}
I_\alpha(t)=e\frac{d}{dt}{\rm}{\rm Tr}\left\{\hat{N}_\alpha \hat{\rho}(t)\right\}, \label{eq:Igen}
\end{equation}
where $\hat{N}_\alpha=\sum_{k\si}c^\dag_{\alpha k\si}c_{\alpha k\si}$ is the occupation-number operator in lead $\alpha$, and  $\hat{\rho}(t)$ is the density matrix of the total system (double dot and leads)  and $e>0$ is the absolute value of the electron charge.
Similarly the {\em heat} current in lead $\alpha$  is given by 
\begin{equation}
J_\alpha(t)=-\frac{d}{dt}{\rm}{\rm Tr}\left\{ \left[\hat{H}_\alpha -\mu_\alpha \hat{N}_\alpha \right]\hat{\rho}(t)\right\} \label{eq:Jgen}
\end{equation}
and it represents the (negative) rate of change of the energy in lead $\alpha$ measured with respect to the local chemical potential.

Performing an adiabatic expansion similar to the one carried out for $\P(t)$, 
these currents can be expressed as a series of contributions of order $\Omega^k$, 
\begin{equation}\label{eq:expansion}
I_{\alpha}(t) = \sum_{k\ge0}I^{(k)}_{\alpha}(t),\qquad J_{\alpha}(t) = \sum_{k\ge0}J^{(k)}_{\alpha}(t). 
\end{equation}
In the limit of weak coupling to the leads  ($\hbar\Gamma\ll k_{\rm B} T$), to lowest order in $\Gamma$ these  are given by\cite{Splett06,Cavaliere09}
\begin{subequations}\label{eq:currents}
\begin{align}
I_\alpha^{(k)}(t) &=  {\bf e}\, \boldsymbol{\mathcal{I}}^{\alpha}_t \boldsymbol{p}^{(k)}_t,\\
J_\alpha^{(k)}(t) &=  {\bf e}\, \boldsymbol{\mathcal{J}}^{\alpha}_t \boldsymbol{p}^{(k)}_t, \label{eq:Jexpansion}
\end{align}
where ${\bf e}=(1,1,1,1,1,1,1,1,1)$ 
and the kernels $\boldsymbol{\mathcal{I}}^{\alpha}_t$ and $\boldsymbol{\mathcal{J}}^{\alpha}_t$ take respectively into account the charge and the heat that flow from lead $\alpha$ into the double dot
\begin{gather}
\left[\boldsymbol{\mathcal{I}}^{\alpha}_t \right]_{ij}=-e(n_i-n_j)\left[ \boldsymbol{\mathcal{W}}_{\alpha,t} \right]_{ij},\\
\left[\boldsymbol{\mathcal{J}}^{\alpha}_t \right]_{ij}=\{(E_i-E_j)-\mu_{\alpha}(n_i-n_j)\}\left[ \boldsymbol{\mathcal{W}}_{\alpha,t} \right]_{ij}.
\end{gather}
\end{subequations}
Here, $E_i$ and $n_i$ are the energy and the number of electrons in the double dot in state $|i\rangle$, respectively.  
The matrix elements of the kernel, $[\W_{\alpha,t}]_{ij}$, represent the probability per unit time that a tunneling event from/to lead $\alpha$ induces the transition $|j\rangle \to |i\rangle$ ($i\neq j$), to lowest order in $\Gamma$. 

The zeroth-order contributions $I_{\alpha}^{(0)}(t)$ and $J_{\alpha}^{(0)}(t)$ represent the steady-state charge and heat currents that would flow in the leads  in a stationary situation with the time-dependent parameters frozen at time $t$. They are non zero only if the system is brought out of equilibrium  by applying a bias voltage or a temperature gradient. 
In the following we will also refer to them as instantaneous currents.

Vice versa, terms with $k>0$ describe the additional contribution to the currents due to the delayed response of the system to the time-dependent modulation.  
From Eqs.~\eqref{eq:currents} and the sum rule $\sum_i[{\W}_{t}]_{ij}=0$, it follows directly that they satisfy the identities
\begin{eqnarray}
 I_{\rm L}^{(k)}(t)+I_{\rm R}^{(k)}(t) & = & -e\frac{d}{dt}\langle \hat{n}\rangle_t^{(k-1)},\label{eq:Itot}\\
 J_{\rm L}^{(k)}(t)+J_{\rm R}^{(k)}(t) & = & \sum_i   E_i(t) \frac{d }{dt}p^{(k-1)}_{i,t}\!-VI^{(k)}(t), \label{eq:Jtot}
 \end{eqnarray}
where $\langle \hat{n}\rangle_t^{(k)}=\sum_i n_i p_{i,t}^{(k)}$ and $I^{(k)}(t) = \frac12(I^{(k)}_\mathrm{R}(t)-I^{(k)}_\mathrm{L}(t))$ are the $k^{\rm th}$ non-adiabatic  corrections to the occupation of the double dot and to the current flowing through it.

The first equation represents essentially the charge continuity equation, and it ensures the conservation  of the charge at every order of the frequency expansion.\footnote{For the contributions to zero-th order in frequency it is $I_{\rm L}^{(0)}+I_{\rm R}^{(0)}=0$, and $J_{\rm L}^{(0)}+J_{\rm R}^{(0)}=-IV^{(0)}$.   }  A direct consequence of Eq.\eqref{eq:Itot} is that the time-averaged electric current is conserved $\bar{I}_{\rm L}^{(k)}=-\bar{I}_{\rm R}^{(k)}$, where the 
bar indicates the average over one driving period
$$
\overline{x}=\frac{\Omega}{2\pi}\int_0^{2\pi/\Omega}dt\, x(t).
$$ 

The second equation expresses instead the first principle of thermodynamics, which relates the total heat flowing into a system to the increase of internal energy of the system itself and the work done by the external sources. This becomes more evident by rewriting the first term on the right hand side of Eq.\eqref{eq:Jtot} in terms of the internal energy of the double dot $\langle E\rangle_{t}^{(k)}=\sum_i E_i p_{i,t}^{(k)}$ and the power delivered by the ac-sources applied to the gates\cite{Solinas13} 
 $\mathcal{P}^{(k)}_\mathrm{ac}=\sum_{\alpha=\mathrm{L,R}}\frac{d \epsilon_\alpha}{dt}\langle \hat{n}_\alpha\rangle_t^{(k-1)}$. 
In this way Eq.\eqref{eq:Jtot} becomes
\begin{equation}
 J_{\rm L}^{(k)}(t)+J_{\rm R}^{(k)}(t) =\frac{d}{dt}\langle E\rangle^{(k-1)}_{t}  -\mathcal{P}^{(k)}_\mathrm{ac}-\mathcal{P}^{(k)}_\mathrm{dc}, \label{eq_work_heat}
\end{equation}
where $\mathcal{P}^{(k)}_\mathrm{dc}=VI^{(k)}$ is the power delivered by the external dc-source according to Joule's law. Note that $\mathcal{P}^{(k)}_\mathrm{dc}$ is positive if the current flows in the direction set by the bias, and negative otherwise (see discussion in Sec.~\ref{sec_efficient}).   
Integrating such an expression over one cycle gives 
\begin{equation}\label{eq:1princ}
 \bar{J}_{\rm L}^{(k)}+\bar{J}_{\rm R}^{(k)} = -(\bar{\mathcal{P}}^{(k)}_\mathrm{ac}+\bar{\mathcal{P}}^{(k)}_\mathrm{dc}), 
\end{equation}
which allows to express the work per cycle done by the ac-sources in terms of the heat currents flowing in the leads.~\cite{Arrachea07,Pekola12b}

\section{Charge and heat pumping }\label{sec_ideal}
The time-dependent modulation of the voltages applied to the local gates  of the quantum dot allows to pump charge and heat through the system. In the following, we will  discuss the transport features of such a double-dot pump and how it can be understood as a nanoscale engine. 
In all the calculations presented below, the pumping cycle is parametrized in terms of the mean energy $E(t)=\bar{E}+\delta_{E}\cos(\Omega t+\phi)$ and the detuning $\e(t)=\bar{\e}+\delta_{\e}\cos(\Omega t)$.

\subsection{Pure adiabatic pumping}\label{sec_quantized}

\begin{figure}
\includegraphics[width=0.9\columnwidth]{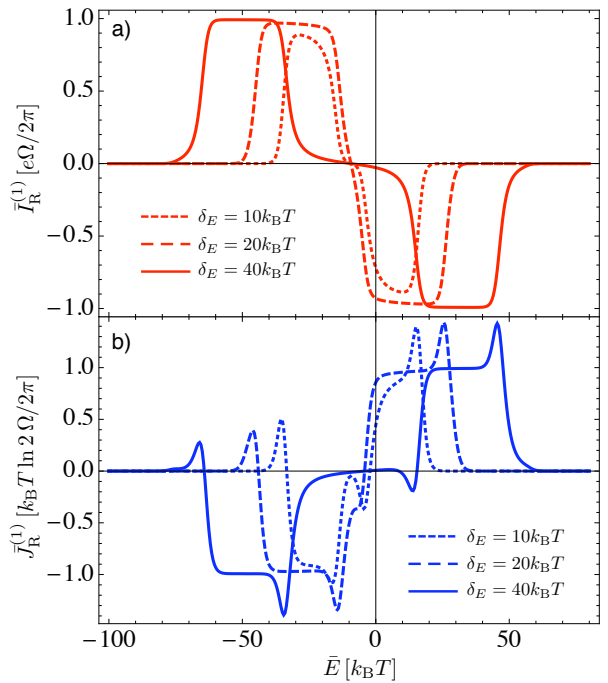}
\caption{(Color online)  Contributions to first order in $\Omega$ to the average charge (a) and heat (b) currents, plotted as a function of the mean energy $\bar{E}$.
The pumping cycle is defined by: $\delta_{\e}=2\delta_{E}$, $\bar{\e}=0$, $\phi=\pi/2$ and $\Omega=\Gamma/200$, and it corresponds to a circular orbit centered around zero detuning.
In both panels:  $T_{\rm L}=T_{\rm R}=T$, $V=0$ and $U=20 k_{\rm B}T$, $t_{\rm c}=10 k_{\rm B}T$, $\Gamma_{\rm L}=\Gamma_{\rm R}=\Gamma/2$, $\hbar\Gamma=k_{\rm B}T/4$.}
\label{fig:currents-omega1-eq}
\end{figure}

To understand the features in  the heat and charge currents due to time-dependent driving, we start by considering the case of pure {\em adiabatic pumping},
meaning that the two leads are at equilibrium ($T_L=T_R=T$ and $V=0$) and transport is only due to the {\em slow} modulation of the levels of the two dots ($\Omega\to0$). 
In this case, the relevant contributions to the currents are  the ones to first order in the driving frequency, $I^{(1)}_\alpha(t)$ and $J^{(1)}_\alpha(t)$.

If the leads are in equilibrium, the charge current pumped through the double dot takes the simple form~\cite{Riwar10}
\begin{eqnarray}\label{eq:I1equi}
I^{(1)}_{\alpha}(t) & = & -e\sum_{\eta=\mathrm{a,b}}\frac{\Gamma_{\alpha,\eta}}{\Gamma_{\eta}}\left(\frac{d}{dt}p^{(0)}_\eta+\frac{d}{dt}p^{(0)}_\mathrm{d}\right) \ ,
\end{eqnarray}
with $p_{\eta}^{(0)}=\sum_{\sigma=\uparrow,\downarrow}p_{\eta\sigma}^{(0)}$, and $p_{\rm d}^{(0)}=\sum_{\sigma,\sigma'=\uparrow,\downarrow}p_{\sigma\sigma'}^{(0)}$, where $p_{i}^{(0)}=e^{-E_{i}/k_{\rm B}T}/(\sum_{i}e^{-E_{i}/k_{\rm B}T})$. It can be divided in two contributions $I_{\alpha}^{(1)}=I_{\alpha,\rm b}^{(1)}+I_{\alpha,\rm a}^{(1)}$, each of which contains only transitions that involve tunneling in or out of the bonding or the antibonding state, respectively. These two terms always contribute to the current with opposite signs\cite{Riwar10} and, while the first is dominant  around the triple point located at $(\epsilon_{L},\epsilon_{R})\approx (0,0)$, the second one is largest around $(\epsilon_{L},\epsilon_{R})\approx (-U,-U)$. This leads to the sign change of the pumped charge as a function of the mean energy $\bar{E}$ shown in Fig.~\ref{fig:currents-omega1-eq}a. Here, we plot the time-averaged charge current flowing into the right lead for various amplitudes of the ac-modulation applied to the gates that forces the state of the system to follow orbits in parameter space similar to those of Fig.~\ref{fig_setup}c.  When the amplitude is large enough so that the pumping cycle fully encircles a triple point, see e.g. the red orbit in Fig.~\ref{fig_setup}c, electrons can be transferred one by one through the double dot, generating a quantized dc-current $\bar{I}=\pm e\Omega/2\pi$.  

One of the key ingredients of such a quantized pumping regime is the alternate decoupling 
from the leads:\cite{Kashcheyevs04}  
whenever one of the two dots comes in resonance with its neighboring lead, the other one is strongly off-resonant, so that particle exchange occurs only with one of the leads at a time. In Eq.\eqref{eq:I1equi}, the ``alternate decoupling'' is encoded in the time-dependent prefactors ${\G_{\alpha,\eta}}/{\G_{\eta}}$ 
 and it can be achieved only if the detuning is much larger than the inter-dot coupling $\e\gg t_{\rm c}$ at the time when the level of one dot crosses the Fermi energy of the neighboring lead, see Eq.\eqref{eq:effective_coupling}.  This in turn requires the modulation amplitude to be larger than $t_{\rm c}$. 
The second ingredient of quantized charge pumping is the strong inter-dot Coulomb interaction $U$, which separates the two triple points well apart and which forbids, for example, the double occupation of the double dot along an orbit that encircles only the triple point around  $(\epsilon_{L},\epsilon_{R})\approx (0,0)$. This permits to transfer charges sequentially from one lead to the other through the two dots according to the direction set by the pumping cycle. Vice versa, no directional transfer is possible if the orbit encircles both triple points, so that the maximal width of the plateaux in the pumped charge current is of the order $U+t_{\rm c}$, further reduced by temperature smearing.

The heat current in the leads has in general a rather  complicated analytic expression, but in the limit in which the occupation of the anti-bonding state can be neglected (i.e. $t_{\rm c}\gg k_{\rm B}T$), it is well approximated by
\begin{equation}\label{eq:J1equi}
J_{\alpha}^{(1)}\approx-\epsilon_{\rm b}\frac{\G_{\alpha,\rm b}}{\G_{\rm b}} \frac{d}{dt}p^{(0)}_{0}+ (\epsilon_{\rm a}+U)\frac{\G_{\alpha,\rm a}}{\G_{\rm a}} \frac{d}{dt}p^{(0)}_\mathrm{d}\, .
\end{equation} 
This equation indicates  that a change in the probability for the double dot to be empty or doubly occupied results in a heat current in the leads that  is directly proportional to the energy involved in the transition responsible for the change. These two contributions are weighted by   different time-dependent prefactors, $\Gamma_{\alpha,\rm b}/\G_{\rm b}$ and $\Gamma_{\alpha,\rm a}/\G_{\rm a}$, which results in the fact that, as for the charge current, changes in $p_{0}^{(0)}$ and $p_{\rm d}^{(0)}$ contribute with different signs to the heat current.

Interestingly, in the regime where the charge current is quantized, also the heat current shows well defined plateaux with height $\pm \frac{\Omega}{2\pi}  k_{\rm B} T \ln 2$, see Fig.~\ref{fig:currents-omega1-eq}b. The emergence of these plateaux can be well understood. In fact, in the regime of quantized charge pumping, $\bar{J}_{\alpha}^{(1)}$
can be directly related to the entropy difference $\Delta S^{(0)}_{\alpha}$ between two charge configurations  that differ only by one electron in dot $\alpha$: 
\begin{equation}\label{eq_Qalpha_approx}
 \frac{2\pi}{\Omega}\bar{J}_{\alpha}^{(1)}= k_{\rm B}T\int_{0}^{2\pi/\Omega}\!\!dt \frac{\G_{\alpha,\eta}}{\G_{\eta}}\frac{d}{dt}S^{(0)}=k_{\rm B}T \Delta S^{(0)}_{\alpha}\, .
\end{equation}
Here, $S^{(0)}=-\sum_{i}p_{i}^{(0)}\ln p_{i}^{(0)} $ is the Shannon entropy of the double dot and $\eta=\rm b\, (a)$ for an orbit fully encircling the triple point around $(\epsilon_{\rm L},\epsilon_{\rm R})\approx (0,0)$ (around $(\epsilon_{\rm L},\epsilon_{\rm R})\approx (-U,-U)$). Along this orbit, the time-dependent prefactor ${\G_{\alpha,\eta}}/{\G_{\eta}}$ equals one close to the resonance with lead $\alpha$ and zero otherwise, so that the integral results in the difference in entropy of the double dot before and after an electron has tunneled through the $\alpha$ barrier, i.e. $\Delta S^{(0)}_{\alpha}=\pm\ln2$. The plus or minus sign corresponds to an electron tunneling in or out of the double dot, according to the direction set by the pumping cycle. 

Note that $k_{\rm B}T \ln 2$ is the minimal amount of energy required to erase one bit of information, according to Landauer's principle.\cite{Landauer61, Bennet82,Maruyama09}  In this case the bit of information is encoded in the spin of the electron in the bonding state, the erasing procedure corresponds to raising the level of one dot
and allowing the electron to tunnel out.\cite{Diana13} The energy $k_{\rm B}T \ln 2$ is provided by the external ac-fields, and results in heat that flows into the lead the electron has tunneled to.
The situation is specular for the reverse process, so that at the end of the cycle the energy $k_{\rm B} T \ln2 $ has been transported from one lead to the other. 

\begin{figure}[tb]
\includegraphics[width=0.95\columnwidth]{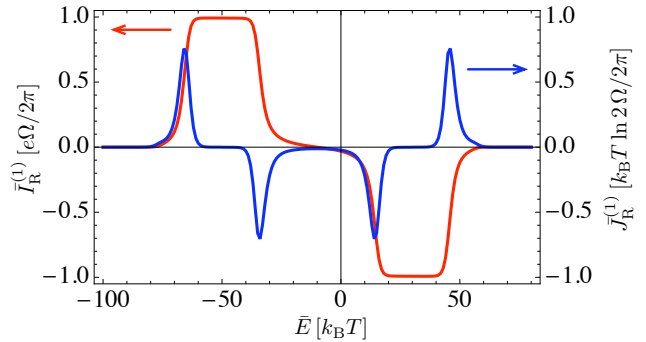}
\caption{Contributions to first order in $\Omega$ to the average charge and heat currents for the case of a fully spin-polarized system. The pumping cycle is defined by: $\delta_{E}=40 k_{\rm B}T$, $\delta_{\e}=2\delta_{E}$, $\bar{\e}=0$, $\phi=\pi/2$ and $\Omega=\Gamma/200$. 
Other parameters:  $T_{\rm L}=T_{\rm R}=T$, $V=0$  and  $U=20 k_{\rm B}T$, $t_{\rm c}=10 k_{\rm B}T$, $\Gamma_{\rm L}=\Gamma_{\rm R}=\Gamma/2$, $\hbar\Gamma=k_{\rm B}T/4$.
\label{fig_spinless}}
\end{figure}

While the appearance of plateaux in the charge current is directly related to the quantization of charge, plateaux in the heat current reflect  specific degeneracies occurring in the system and are therefore tunable e.g. by an external magnetic field. This is shown in Fig.~\ref{fig_spinless}, where the plateaux in the heat current are fully suppressed by a strong magnetic field that spin-polarizes the system $B\gg k_\mathrm{B}T$. Vice versa those in the charge current are unaffected. In this case, charge transport is not accompanied by heat transfer from one lead to the other. From the information-theory point of view, one could say that the regular flow of spinless electrons produced by the double-dot pump in a high magnetic field does not carry any useful information and, therefore, it is not accompanied by the heat transfer which otherwise would be required by Landauer's principle.    

Another feature that distinguishes heat from charge pumping is the appearance of peaked shaped features at the borders of the plateaux. These peaks approach the values  $\pm\frac{\Omega}{2\pi}k_{\rm B} T\ln 3$ and $\pm \frac{\Omega}{2\pi}k_{\rm B} T\ln \frac{3}{2}$ and are associated to orbits that touch one of the triple points, see e.g. the black orbit in Fig.~\ref{fig_setup}c. In this case particles are also exchanged with the leads in situations in which two different charge states (e.g. singly and doubly occupied) are degenerate, and this results in differences of the entropy between initial and final state of a certain loading and unloading process that equal $\pm \ln 3$ or $\pm \ln \frac32$.  
This change in entropy does not translate entirely in the heat exchanged with one lead  because close to a triple point the energy dependence of the weight factor ${\G_{\alpha,\eta}}/{\G_{\eta}}$ in Eq.\eqref{eq_Qalpha_approx}, cannot be neglected or, equivalently, the ``decoupling approximation'' breaks down and each dot exchanges heat with both leads simultaneously.\footnote{ The maximum value $\pm\frac{\Omega}{2\pi}k_{\rm B} T\ln 3$ could be reached in the limit $t_{\rm c}\to 0$, in which the two dots are decoupled. In this case, no particle can flow through the system, but still heat can be transferred from one lead to the other during a pumping cycle, thanks to the strong Coulomb repulsion between electrons in the two dots, see also Ref.\onlinecite{Ruokola11}}

Finally, we notice that if the leads  are in equilibrium Eq.\eqref{eq:Jtot} reduces to
\begin{align}\label{eq:IEtot1}
J_{\rm L}^{(1)}(t)+J_{\rm R}^{(1)}(t)=k_{\rm B}T\frac{d}{d t}S^{(0)}.\end{align}
This expression is equivalent to the Clausius equality of equilibrium thermodynamics 
and it indicates that to first order in $\Omega$ the dynamics of the system is \textit{reversible}  if the leads are in equilibrium. A direct consequence of Eq.\eqref{eq:IEtot1} is that on average, during a cycle, heat is transported from one lead to the other $\bar{J}_{\rm L}^{(1)}=-\bar{J}_{\rm R}^{(1)}$.

\subsection{Pumping against a gradient}\label{sec_gradient}
So far we consider only the {\em case} of pure pumping, in which the leads are in equilibrium and transport is solely due to periodic modulation of dots' levels. However, an important feature of the double-dot pump  is the possibility of achieving quantized charge pumping even in the presence of a finite bias voltage $V$.
From the experimental point of view, pumping against a  bias represents  a clear proof that the measured current is set by the chosen pumping cycle.\cite{Pothier92} Moreover, it also allows  thinking of applications of the double-dot pump e.g. as  ``battery charger'' transferring electrons from a lower to a higher chemical potential.  This will be discussed in the following Section~\ref{sec_efficient}. 

Pumping against a voltage bias requires an orbit in parameter space that minimizes the contributions of the instantaneous currents $\bar{I}^{(0)}_{\alpha}$ and $\bar{J}^{(0)}_{\alpha}$, which -- flowing in the direction set by the gradient irrespectively of the orientation of the pumping cycle -- play the role of leakage currents.
This can be achieved by choosing a pumping cycle that fully encircles a single triple point. However, it has to be taken into account that the triple points, i.e. the regions in parameter spaces where no charge configuration is stable, get broadened by the gradient itself and will eventually merge into a single one for very large  $V$. This poses an upper limit to the maximal voltage against which it is possible to pump, see Sec.~\ref{sec_leakage}.
Nevertheless, as long as the triple point regions are well separate from each other and can be encircled by a controlled modulation of the gates, it remains possible to pump one electron per cycle through the double dot.

\begin{figure}[tb]
\includegraphics[width=3.in]{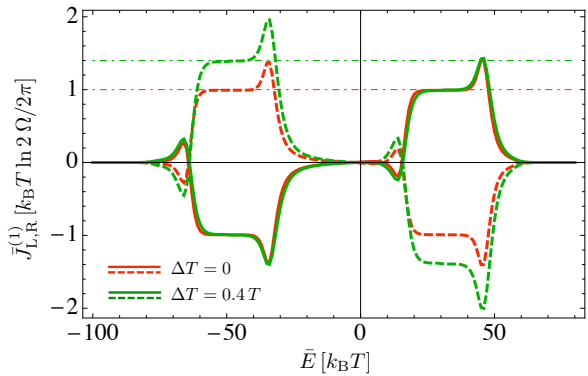}
\caption{(Color online) Contributions to first order in $\Omega$ to the average heat current in the left (dashed lines) and the right lead (full lines), with and without a temperature gradient. Here we assumed $T_{\rm L}=T+\Delta T$ and $T_{\rm R}=T$, with $\Delta T=0$ (red lines) or $\Delta T=0.4T$ (green lines).  The thin dot-dashed lines correspond to $k_{B}(T+\Delta T)\ln2\, \Omega/2\pi$. 
The heat current in each lead shows plateaux proportional to the local temperature $T_{\alpha}$. When $\bar{J}^{(1)}_{\rm L}<0$ and $\bar{J}^{(1)}_{\rm R}>0$, heat is pumped form the cold to the hot reservoir .
The pumping cycle is defined by: $\delta_{E}=40 k_{\rm B}T$, $\delta_{\e}=2\delta_{E}$, $\bar{\e}=0$, $\phi=\pi/2$ and $\Omega=\Gamma/200$. 
Other parameters:   $U=20 k_{\rm B}T$, $t_{\rm c}=10 k_{\rm B}T$, $\Gamma_{\rm L}=\Gamma_{\rm R}=\Gamma/2$, $\hbar\Gamma=k_{\rm B}T/4$.
\label{fig_tempgrad}}
\end{figure}

The situation is similar for what concerns  heat pumping when the leads have different temperatures: as long as it is possible to choose a pumping cycle that fully encloses a triple point, the heat currents in each lead exhibits well defined plateaux of height $\frac{\Omega}{2\pi} k_{\rm B}T_{\alpha}\ln 2 $,  see Fig.~\ref{fig_tempgrad}. In other words, even if the system is globally brought out of equilibrium, the heat exchanged singularly with each lead obeys Clausius relation $\bar{J}_{\alpha}^{(1)}=k_{\rm B}T_{\alpha} \Delta S^{(0)}_{\alpha}\frac{\Omega}{2\pi}$.
This is again a consequence of the ``alternate decoupling'' occurring along an orbit fully encircling a triple point, which ensures that the double dot is temporarily coupled only to one lead at a time.  In the considered limit of slow driving $\Omega \to 0$,  the double dot has time to equilibrate with the lead it is coupled to. As a consequence,  even in the presence of a temperature gradient, processes that change the total occupation of the double dot represent isothermal transitions between equilibrium states.

In the limit of slow driving,  a pumping cycle that fully encloses a triple point can then be regarded as  a 
nanoscale analog of the Carnot cycle in which two isothermal transitions are connected by two adiabatic ones (where now adiabatic means that the entropy of the system remains unchanged). The isothermal processes are those that change the total occupation of the double dot, and the adiabatic ones 
 those that accompany the crossing of the two dots' levels, which, occurring when both dots are far away from the Fermi energy of the leads,  represent reversible processes. 

Finally we notice that 
 the total average heat-current flowing into the system $\bar{J}_{\rm L}^{(1)}+\bar{J}_{\rm R}^{(1)}$ is directly related to the excess entropy production due to the time-dependent driving.\cite{Sagawa11} As emphasized in Ref.\onlinecite{Sagawa11}, this is a geometric quantity that depends only on the trajectory in parameter space. The geometric nature of the excess entropy production -- and more in general of the pumped charge and heat to first order in frequency -- is a direct consequence of the relation between $\bar{I}_{\alpha}^{(1)}$ and  $\bar{J}_{\alpha}^{(1)}$ and the time derivative of the instantaneous occupation probability $\P^{(0)}$, see Appendix.~\ref{app_geometric}.  \footnote{See also Ref.\onlinecite{Ren10}, where heat pumping in a molecular junction connected to bosonic baths is considered.}

\subsection{Operation of a double-dot device}\label{sec_efficient}

The possibility of pumping charge and/or heat against a gradient 
shows that a driven double-dot device can have different applications as a \textit{nanoscale engine}. In the following we discuss in detail three different operating modes: the {\em charge pump} moving electrons against a bias voltage,  the {\em heat pump} transferring heat from a cold to a hot reservoir, and the {\em thermal engine} producing work extracting energy from a hot reservoir and releasing part of it into a cold one. To quantify their performance, we introduce efficiencies in analogy to the classical counterparts. 
In this section we concentrate on the {\em ideal} working regime where the main contributions to the pumped currents  are of first order in the driving frequency, i.e. $\bar{I}_{\alpha}\approx\bar{I}_{\alpha}^{(1)} $ and $\bar{J}_{\alpha}\approx\bar{J}_{\alpha}^{(1)} $. Corrections due to finite operation-time and leakage currents will be analyzed in Section~\ref{sec_limitations}.

\subsubsection{Charge pump}

When the double dot is operated in a way as to transfer electrons from a lower to a higher chemical potential, it can be regarded as a ``battery charger'' doing work on the dc-source.  We characterize the performance of such an engine by its energy conversion efficiency
\begin{equation} \label{eq:eta_ch}
\eta_{\rm ch}=\frac{-\bar{I}V}{\bar{\mathcal{P}}_{\rm ac}},
\end{equation}
where $-\bar{I}V$ is the useful work per unit time done by the pump on the dc-voltage source, and $\bar{\mathcal{P}}_{\rm ac}$ is the average power delivered by the ac-fields applied to the gate electrodes. 
The sign convention used here for the current is that $\bar{I}$ is positive when it flows in the direction set by the dc source, so that $-\bar{I}V>0$ when the 
 driven quantum dot  actually works as a {\em charge pump} moving electrons from a lower to a higher chemical potential. Only in this case it make sense to speak about  useful work done by the pump and to characterize its efficiency in terms of Eq.\eqref{eq:eta_ch}.
 
The power delivered by the ac-fields $\bar{\mathcal{P}}_{\rm ac}$ can be quantified in terms of the heat currents flowing into the leads according to Eq.\eqref{eq:1princ}, i.e. $\bar{\mathcal{P}}_{\rm ac}=-\bar{\mathcal{P}}_{\rm dc}-\sum_{\alpha}\bar{J}_\alpha$, where $\bar{\mathcal{P}}_{\rm dc}=\bar{I}V$ is the work per unit time done  {\em by} the dc-source. The efficiency of the double-dot pump as a battery charger is then given by  
\begin{equation}\label{eq_eff_ch}
\eta_\mathrm{ch} = \frac{\bar{I}V}{\bar{J}_L+\bar{J}_R+\bar{I}V}.
\end{equation}

If the leads have the same temperature, the total heat current is zero along an orbit that fully encloses a triple point and the efficiency  takes its maximal value $\eta_\mathrm{ch} =1$. In this case, the work done by the source is fully used to transfer charge against a potential.

\subsubsection{Heat pump}
As discussed in Sec.~\ref{sec_gradient}, a double dot operating between two reservoirs with different temperatures along an orbit that fully encloses a triple point, represents an analog of the Carnot engine. Depending on the direction of the cycle, the device acts  either as a refrigerator or as a heat engine.  
 We are interested here in the situation where no voltage bias is present. 
 
The double dot is operated as a refrigerator when heat is transferred from a cold to a hot reservoir during the pumping cycle. 
The efficiency of a  heat pump  
is in general characterized by the coefficient of performance 
\begin{equation} \label{eq:eta_ht}
{\rm COP}=\frac{\bar{J}_{\rm cold}}{\bar{\mathcal{P}}_{\rm ac}},
\end{equation}
where $\bar{J}_{\rm cold}$ is the average heat flow {\em out of} the cold reservoir 
and $\bar{\mathcal{P}}_{\rm ac}$ the power delivered by the ac-field during one cycle. 
If no bias is applied, the latter is directly related to the total heat flowing out of the double dot $\bar{\mathcal{P}}_{\rm ac}=-\sum_{\alpha}\bar{J}_{\alpha}$, see Eq.\eqref{eq:1princ}.  Along an orbit that fully encircles a triple point, the heat current in each lead is directly proportional to the temperature of that  lead. In this case $\bar{J}_{\rm cold}=\frac{\Omega}{2\pi}k_{\rm B}T_{\rm cold}\ln 2$ and $\bar{\mathcal{P}}_{\rm ac}=\frac{\Omega}{2\pi}k_{\rm B}(T_{\rm hot}-T_{\rm cold})\ln 2$ and the cooling efficiency of the double-dot pump reaches the Carnot limit ${\rm COP}_{\rm Carnot}= T_\mathrm{cold}/(T_\mathrm{hot}-T_\mathrm{cold}),
$
confirming the analogy between the ``ideal'' pumping cycle and the Carnot cycle.

\subsubsection{Heat engine}
The double-dot pump acts as a heat engine producing work on the external fields, when the pumping cycle is such that heat is extracted from the hot reservoir and released into the cold one.  For the case of Fig.~\ref{fig_tempgrad}, this corresponds to the regimes where $\bar{J}^{(1)}_{\rm L}>0$ and $\bar{J}^{(1)}_{\rm R}<0$.
The performance of a heat engine  is characterized by the efficiency coefficient
\begin{equation} \label{eq:eta_te}
\eta=\frac{-\bar{\mathcal{P}}_{\rm ac}}{\bar{J}_{\rm hot}}.
\end{equation}
where  $-\bar{\mathcal{P}}_{\rm ac}$ is now the work per unit time \textit{done by the system on the ac fields} and  $\bar{J}_{\rm hot}$ is the heat current absorbed from the hot reservoir.

Similarly to the case of the refrigerator, along an orbit that fully encircles a triple point, we find $-\bar{\mathcal{P}}_{\rm ac}=\frac{\Omega}{2\pi}k_{\rm B}(T_{\rm hot}-T_{\rm cold})\ln 2$ and $\bar{J}_{\rm hot}=\frac{\Omega}{2\pi}k_{\rm B}T_{\rm hot}\ln2$, and  
the efficiency of the double-dot pump reaches again the Carnot limit $\eta_{\rm Carnot}=1-{T_{\rm cold}}/{T_{\rm hot}}
$.

\section{Limitations of a realistic pump}\label{sec_limitations}

In the previous section we have assumed that the charge and heat currents 
can be approximated solely by the first non-adiabatic corrections,  $\bar{I}_{\alpha}\approx\bar{I}_{\alpha}^{(1)} $ and $\bar{J}_{\alpha}\approx\bar{J}_{\alpha}^{(1)} $.  This correspond to considering the case of extremely slow driving $\Omega\to0$ and to implicitly assuming that the instantaneous currents  $\bar{I}_{\alpha}^{(0)} $ and $\bar{J}_{\alpha}^{(0)}$ can be neglected along an orbit that fully encircles a triple point. In this ideal case, the efficiencies of the double-dot based engines discussed in Sec.~\ref{sec_efficient},  reach their maximum  theoretical values.  In this section we investigate instead the limitations to the performance of a realistic double-dot pump that  are due to  a small but finite driving frequency and to leakage currents.

\subsection{Corrections due to finite operation-time}\label{sec_heating}

\begin{figure}
\includegraphics[width=0.9\columnwidth]{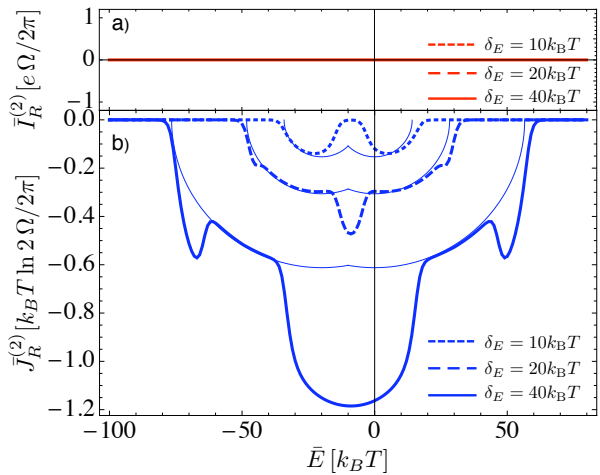}
\caption{(Color online) Contributions to second order in $\Omega$ to the average charge (a) and heat (b) currents, plotted as a function of the mean energy $\bar{E}$.
The pumping cycle is defined by: $\delta_{\e}=2\delta_{E}$, $\bar{\e}=0$, $\phi=\pi/2$ and $\Omega=\Gamma/200$.
The thin lines in panel (b) represent the approximate expression $\bar{J}_{\alpha}^{(2)}\approx- \frac{3}2\frac{v_{\alpha}}{\Gamma}\frac{\Omega}{2\pi}$.
In both panels:  $T_{\rm L}=T_{\rm R}=T$, $V=0$ and $U=20 k_{\rm B}T$, $t_{\rm c}=10 k_{\rm B}T$, $\Gamma_{\rm L}=\Gamma_{\rm R}=\Gamma/2$, $\hbar\Gamma=k_{\rm B}T/4$.
\label{fig:currents-omega2-eq}}
\end{figure}

To take into account the effects of a finite driving frequency, we evaluate  the second non-adiabatic corrections to the charge and the heat currents, i.e. the 
contributions to second order in the frequency,  ${I}_{\alpha}^{(2)}$, ${J}_{\alpha}^{(2)}$.  For simplicity, we start again by first discussing the case where the leads are at equilibrium and postpone the case of pumping against a gradient to Sec.~\ref{sec_leakage}. 
 Moreover, we focus on the regime of large modulation amplitudes, which is the most interesting for applications. The  opposite case of small amplitudes pumping is described in detail in the Appendix.~\ref{app_secondorder}.

 In Fig.~\ref{fig:currents-omega2-eq} we plot the contributions to second order in $\Omega$ to the average charge and heat currents $\bar{I}_{\rm R}^{(2)}$, $\bar{J}_{\rm R}^{(2)}$, for the same choice of parameters considered in Fig.~\ref{fig:currents-omega1-eq}.   Comparing these figures it is apparent that, even in the case of relatively slow driving $\Omega=\Gamma/200$ considered here, the contribution to second order in $\Omega$ to the heat current represents a significant amount of the total heat exchanged with a lead per cycle. Vice versa, the contributions to second order in $\Omega$ to the charge current can  be completely neglected. 
 
This behavior is easily understood considering again an orbit in parameter space that fully encloses a triple point. As discussed before, along such an orbit the exchange of particles and heat with the leads involves only one dot and its neighboring lead at a time, while the other lead is temporarily decoupled due to the large inter-dot detuning. Particle exchange occurs when the level of a dot crosses the Fermi energy of the neighboring lead. However, if the level moves with a finite velocity, electrons can tunnel  out of (in) the dot not only in resonance, but also at a somewhat higher (lower) energy,  resulting the emission of in a hot electron (hole) into the leads.  Both processes contribute to increasing the energy of the lead and give rise to a negative heat current (meaning that it is flowing from the dot into the leads). In other words, $\bar{J}^{(2)}_{\alpha}$ represents the irreversible heat production that accompanies the action of the ac-fields applied to the double dot. 
The maximal excess energy that can be deposited into the leads is determined by the  amplitude of the modulation $\delta_{E}$,  while the probability that strongly off-resonant transition occurs is given by the ratio $\Omega/\Gamma$, so that the heat current to second order in $\Omega$ scales as $|\bar{J}_{\alpha}^{(2)}|\sim \delta_{E} \Omega^{2}/\Gamma$.   A more accurate estimate leads to 
\begin{equation}\label{eq:approx}
\bar{J}_{\alpha}^{(2)}\approx-\frac{3}2\frac{v_{\alpha}}{\Gamma}\frac{\Omega}{2\pi},
\end{equation}
where $v_{\alpha}$ is the speed 
with which the level of dot $\alpha$ crosses the Fermi energy of lead $\alpha$.
For a circular orbit fully encircling the triple point around $(\epsilon_{\rm L},\epsilon_{\rm R})=(0,0)$ we have $v_{\alpha}=\Omega\sqrt{2}\delta_{E}\sqrt{1-{\bar{\epsilon}_{\alpha}^{2}}/{2\delta_{E}^{2}}}$.\footnote{This expression holds only for large orbits centered around $(\epsilon_{\rm L},\epsilon_{\rm R})=(0,0)$, for which it is $\bar{\epsilon}_{\alpha}^{2}/2\delta_{E}^{2}<1$} 
 For an orbit around the other triple point, $\bar{\epsilon}_{\alpha}$ should be replaced by $(\bar{\epsilon}_{\alpha}+U)$ in the expression for $v_{\alpha}$. Deviations from the behavior predicted by Eq.\eqref{eq:approx} occur for orbits that cross a  triple point and are associated to an increased heat production close to the turning points of the ac-modulation.
The large peak around $\bar{E}=-U/2$ for large amplitudes corresponds to orbits that enclose both triple points. In this case a dot level crosses the Fermi energy of the neighboring leads twice during each cycle, leading to (almost) a doubling of the heat current $J_{\alpha}^{(2)}$.

It is important to stress that the contributions to second order in $\Omega$ to the heat current are negative in both leads (with $\bar{J}_{\rm L}^{(2)}=\bar{J}_{\rm R}^{(2)}$ in the case of zero average detuning $\bar{\e}=0$),\cite{Wei04} meaning that heat is \textit{deposited in the leads} in every pumping cycle. This pose some serious constrains to the possibility of using a double dot as a heat pump to extract heat from a reservoir, as it requires to reduce significantly the driving frequency $\Omega\ll \ln2\,  \Gamma k_{\rm B} T/{\delta_{E}}$ in order to minimize heating effects. 

\begin{figure}
\includegraphics[width=3.in]{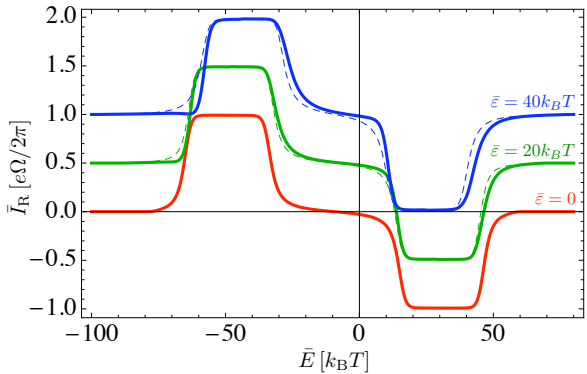}
\caption{(Color online) Average charge current as a function of the mean dot-level energy $\bar{E}$ for different values of the average detuning $\bar{\varepsilon}$. The thick lines represent the total charge current $\bar{I}_{R}=\bar{I}_{R}^{(1)}+\bar{I}_{R}^{(2)}$, while the thin-dashed ones correspond to the contribution to first order in frequency $\bar{I}_{R}^{(1)}$ alone.  
The pumping cycle is defined by: $\delta_{E}=40k_{\rm B}T$, $\delta_{\e}=2\delta_{E}$, $\phi=\pi/2$ and $\Omega=\Gamma/200$.
Other parameters:  $T_{\rm L}=T_{\rm R}=T$, $V=0$  and  $U=20 k_{\rm B}T$, $t_{\rm c}=10 k_{\rm B}T$, $\Gamma_{\rm L}=\Gamma_{\rm R}=\Gamma/2$, $\hbar\Gamma=k_{\rm B}T/4$. Curves are offset by $\frac12e\Omega/2\pi$ for clarity.
\label{fig_charge_secondorder}}
\end{figure}

While the emission of a hot hole or a hot electron contribute with the same sign to the heat current into the leads, the two processes compensate each other in the average charge current, so that  $\bar{I}_{\alpha}^{(2)}$ remains in general pretty small as long as $\Omega\ll\Gamma$.   In particular, it can be shown that $\bar{I}_{\alpha}^{(2)}=0$ for orbits with average zero detuning, thanks to the symmetry of the pumping cycle, see Fig.~\ref{fig:currents-omega2-eq}a. Even for orbits with 
 $\bar{\e}\neq0$, the corrections due to $\bar{I}_{\alpha}^{(2)}$ to the charge current are negligible as long as the pumping cycle fully encircles a triple point, and become sizable only at the edges of the current plateaux, see Fig.~\ref{fig_charge_secondorder}.

Finally, we would like to comment briefly on the consistency of our {adiabatic expansion}. As explained in Sec.\ref{sec:kineticeq}, our calculations are based on the expansion ${\bf p}(t)=\sum_{k}{\bf p}_{t}^{(k)}$, where the contributions ${\bf p}_{t}^{(k)}$ scales as $(\Omega/\Gamma)^{k}$ giving a convergent series in the limit $\Omega\sim\Gamma$.\cite{Splett06,Cavaliere09} The scaling of the various contributions to the average charge and  heat currents  is less obvious, as shown by $\bar{J}_{\alpha}^{(2)}\gtrsim \bar{J}_{\alpha}^{(1)}$ in certain frequency regimes. This is  because the physical processes responsible for the heating induced by the ac-driving scale as $\Omega^{2}$ 
and therefore start contributing to the heat current only from second order, in terms of the expansion we use.
However, it can be shown rigorously that all the contributions $\bar{J}_{\alpha}^{(k\ge2)}$ are suppressed at least by a factor $\Omega/\Gamma$ with respect to $\bar{J}_{\alpha}^{(2)}$, so that it is meaningful to truncate the expansion in Eq.\eqref{eq:Jexpansion} to second order only.

\subsection{Leakage currents \& efficiency of a realistic pump}\label{sec_leakage}
The leakage currents $I_{\alpha}^{(0)}$ and $J_{\alpha}^{(0)}$ pose another important limitation to the operation of the double dot as a heat or charge pump. 
In fact,  it is possible to transfer electrons against a bias only as long as the pumped current $\bar{I}_{\alpha}^{(1)}+\bar{I}_{\alpha}^{(2)}$, i.e. the current resulting from the time-dependent modulation of the dots' levels,  prevails over the instantaneous one $\bar{I}_{\alpha}^{(0)}$. Similarly,  to pump heat from a cold to a hot reservoir it is necessary that the magnitude of  $\bar{J}_{\alpha}^{(1)}+\bar{J}_{\alpha}^{(2)}$ is larger than the one of $\bar{J}_{\alpha}^{(0)}$. The contributions from $I_{\alpha}^{(0)}$ and $J_{\alpha}^{(0)}$ can be strongly suppressed by choosing an orbit along which the levels of the  two dots cross each other only well outside of the bias or temperature window. 
However, for a given  orbit,  the maximal amount of charge or heat that can be pumped in or out of a reservoir in one cycle is fixed, while the leakage currents increase with $V$ and $\Delta T$. This poses an upper limit to the maximum voltage or temperature gradient against which it is possible to pump. 
Moreover, the  contribution of $\bar{I}^{(0)}_{\alpha}$ and  $\bar{J}^{(0)}_{\alpha}$  to the total charge and heat extracted from a lead during a pumping cycle grows as $\Omega^{-1}$ when 
the driving frequency is reduced.  As a consequence, the maximum voltage or temperature gradient  that the double-dot pump can overcome depends both on the chosen pumping orbit and on the speed at which it is traversed.

\begin{figure}
\includegraphics[width=3.in]{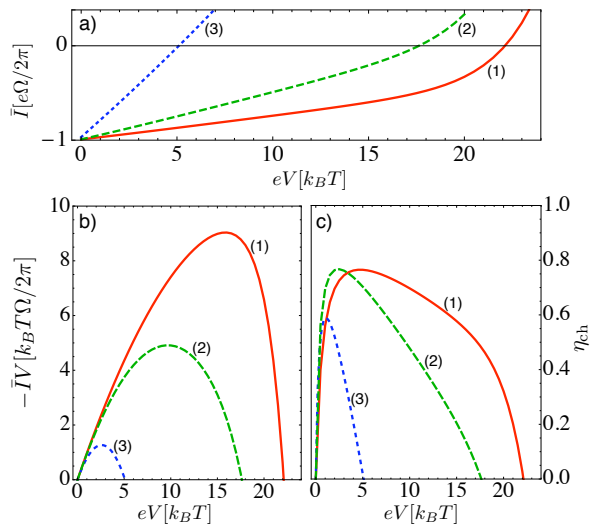}
\caption{(Color online) (a)  Average charge current flowing through the double dot in the presence of a bias voltage and of time-dependent driving. The current is plotted as a function of the applied bias $V$ for three different sets of pumping parameters: 
Curves (1) and (2) correspond to the orbit described by $\delta_E=40 k_{\rm B}T$, $\bar{E}=30 k_{\rm B}T$, $\bar{\e}=0$, $\phi=\pi/2$, traversed with frequency $\Omega=\Gamma/200$ and $\Omega=\Gamma/400$, respectively. Curve (3) correspond to a smaller orbit with $\delta_E=20 k_{\rm B}T$, $\bar{E}=15 k_{\rm B}T$, $\bar{\e}=0$,  $\phi=\pi/2$ and $\Omega=\Gamma/200$.  
 Charge is pumped against the bias as long as $\bar{I}<0$.
(b) Work per unit time done on the dc-source and (c) efficiency of the double-dot pump for the same pumping cycles considered in panel (a). The efficiency is defined according to Eq.\eqref{eq_eff_ch}.  
In all panels: $U=20 k_{\rm B}T$, $t_{\rm c}=10 k_{\rm B}T$, $\Gamma_{\rm L}=\Gamma_{\rm R}=\Gamma/2$, $\hbar\Gamma=k_{\rm B}T/4$ and $T_{\rm L}=T_{\rm R}=T$. 
\label{fig_againstbias}}
\end{figure}

The competition between pumped and leakage currents can be seen e.g. in Fig.~\ref{fig_againstbias}a, for the case of a charge pump working against a bias voltage. Here we show the total charge current $\bar{I}=\bar{I}^{(0)}+\bar{I}^{(1)}+\bar{I}^{(2)}$ flowing through the dot as a function of the applied bias for different amplitudes of the driving cycle and different driving frequencies. In each case we find $\bar{I}^{(2)}\approx0$. 
According to our convention on the sign of $I$,  the pump moves electrons against the direction set by the bias as long as $\bar{I}<0$.  The maximal voltage $V_{\rm max}$ that the pump can sustain is then defined by the condition $\bar{I}=0$.   For larger voltages, the pumped current is overrun by the one due to the bias. As expected, $V_{\rm max}$ is sensibly reduced if the radius of the orbit is shrunk, or if the pumping frequency is lowered.
 
As long as $\bar{I}<0$, the pump performs a positive work per pumping-period, by transferring electrons from a lower to a higher chemical potential,  see Fig.~\ref{fig_againstbias}b. The efficiency of such a pump can be quantified by the energy conversion coefficient $\eta_{\rm ch}$, Eq.\eqref{eq_eff_ch}, where now all contributions up to second order in $\Omega$ to the charge and heat currents have to be taken into account, see Fig.~\ref{fig_againstbias}c. At large voltages,  the major limitations to the performance of the pump come from the leakage current $\bar{I}^{(0)}$, which poses an upper limit to the maximal power that can be delivered by the pump on the dc-source. Vice versa, at low bias the efficiency $\eta_{\rm ch}$ is mostly affected by the heat production term  $\bar{J}_{\rm L}^{(2)}+\bar{J}_{\rm R}^{(2)}$, which represents  the minimal power that the ac-sources have to provide in order to drive the system along the considered orbit. Since this is finite even in the limit of no applied bias, the efficiency of the double-dot charge pump vanishes for $V\to 0$. Interestingly, despite the detrimental effects due to heat production and leakage currents, the double-dot charge pump can reach efficiencies up to 70\% of the ideal value, when operated against a finite voltage along a sufficiently large pumping cycle, see Fig.~\ref{fig_againstbias}c.

\begin{figure}
\includegraphics[width=3.in]{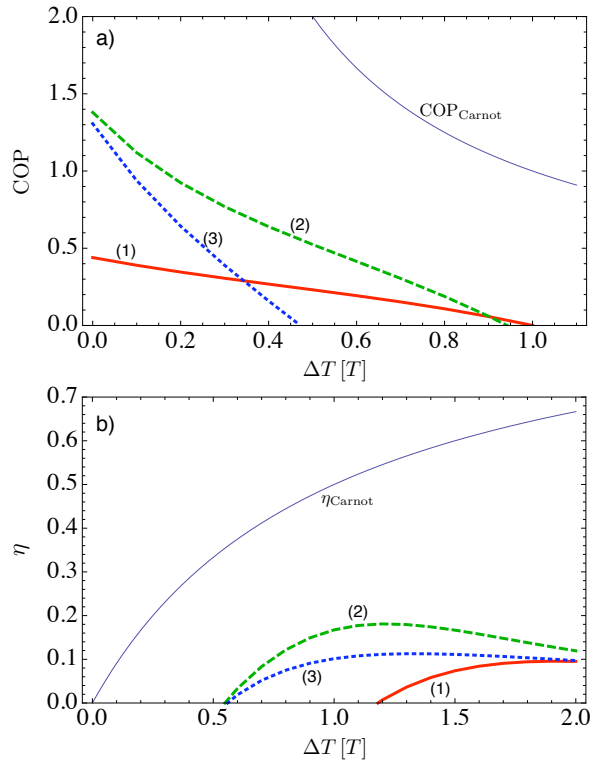}
\caption{(Color online) (a) Coefficient of performance of the double-dot pump as a heat pump, as defined in Eq.\eqref{eq:eta_ht}. In this case we choose $T_{\rm L}=T_{\rm R}+\Delta T$ and a pumping cycle such that heat is transfered from right to left, i.e. against the temperature gradient. The curves labeled as (1)-(3) correspond to the pumping cycles discussed in Fig.~\ref{fig_againstbias}. The COP of an ideal Carnot refrigerator  is also plotted for comparison.
(b) Efficiency of the double-dot pump as a heat engine, as defined in Eq.\eqref{eq:eta_te}. Also in this case $T_{\rm L}=T_{\rm R}+\Delta T$, but the pumping cycle is now such that heat is transfered from left to right, i.e. according to the temperature gradient. The curves labeled as (1)-(3) corresponds again to the  orbits discussed in Fig.~\ref{fig_againstbias}, but now traversed in the opposite direction, i.e. $\phi=-\pi/2$. The efficiency of an ideal Carnot engine is  plotted for comparison.  
In all panels: $U=20 k_{\rm B}T$, $t_{\rm c}=10 k_{\rm B}T$, $\Gamma_{\rm L}=\Gamma_{\rm R}=\Gamma/2$, $\hbar\Gamma=k_{\rm B}T/4$ and $\mu_{\rm L}=\mu_{\rm R}$. 
\label{fig_efficiency_cool}}
\end{figure}

The limitations due to leakage currents and finite operation-time are more severe when the double dot is employed as a refrigerator to pump heat from a cold to a hot reservoir.  In fact, while 
the second non-adiabatic corrections
are mostly negligible for what concerns the charge current, they represent a significant amount of the heat  flowing into the leads. 
Moreover, being always injected from the double dot into the leads, they reduce the possibility of extracting heat from a reservoir.
Suppressing $\bar{J}^{(2)}_{\alpha}$ requires to lower considerably the driving frequency, but this in turn makes the system more sensitive to leakage currents. Both effects reduce significantly the efficiency of the double-dot pump with respect to the Carnot limit.
This can be seen in Fig.~\ref{fig_efficiency_cool}a, where we plot the coefficient of performance of a double-dot heat pump, plotted as a function of the temperature gradient between the leads for the same pumping cycles considered in Fig.~\ref{fig_againstbias}.
The Carnot efficiency is also plotted for comparison.
The effects due to finite-time operation are particularly evident in the limit $\Delta T\to 0$, where an ideal pump would be ``infinitely efficient'', since it require no power to  transfer adiabatically heat from one lead to the other when these have the same temperature.  Vice versa, due to the heating of the leads which accompanies the 
modulation of the levels of the double dot, the power needed to operate a pump at finite speed remains finite, as well as its efficiency.  Moreover, an ideal heat pump would be able to work against an arbitrary temperature gradient, but this is not the case of a real double-dot pump, where the pumped current is  overrun by leakage currents as soon as the temperature gradient $\Delta T$ becomes larger than a critical value.

Heat production limits also the operation of a double-dot pump as a heat engine, see Fig.~\ref{fig_efficiency_cool}b. In fact, while the maximal amount of work that the pump can perform extracting energy from a hot reservoir and releasing it into  a cold one is proportional to the temperature difference, the energy cost of moving along an orbit in parameter space -- which is represented by the heat dissipated into the leads -- is roughly independent of $\Delta T$. This means that if the leads have similar temperatures, the work put into the system by the external ac-fields exceeds the one that can be extracted form the heat reservoirs, leading to a negative energy balance. In this case, the pump does no work and it make no sense to speak about its efficiency in terms of Eq.\eqref{eq_eff_ch}.  Dissipation due to heating  can be reduced by driving the system at a lower frequency, but at the cost of  increasing the effects of the leakage currents. These drag a substantial amount of heat out of the hot reservoir without contributing to the work performed by the system  (since $\bar{J}_{\rm L}^{(0)}+\bar{J}_{\rm R}^{(0)}=0$), suppressing the efficiency of the pump for large $\Delta T$. 

Finally we would like to mention that, apart for leakage currents and heat production, other effects, such as coupling to phonons, cotunneling and missed inter-dots transitions, may affect the performance of the double-dot pump.  However, the last two mechanisms are 
suppressed in the limit of weak coupling to the leads, strong inter-dot hybridization and slow driving frequencies considered in this paper, and coupling to phonons is also expected to be significantly reduced at the cryogenic temperatures at which pumping experiments are performed.

\section{Conclusions}
We investigated charge and heat transport in a driven double-dot device.
We showed that in the regime of quantized charge pumping, i.e. when electrons are transferred one by one through the system thanks to a controlled modulation of the energy levels of the two dots, the heat current exhibits well defined plateaux if the driving frequency $\Omega$ is sufficiently small.
The height of these plateaux is directly proportional to the temperature of the leads and reflects specific degeneracies of the double-dot states involved into transport, namely the spin degeneracy of the  bonding state. In the limit of slow driving  $\Omega\to 0$, the heat current through the double dot can then be controlled by an external magnetic field and be completely suppressed by fully spin polarizing the system. 

While the quantization of the transported charge is rather robust with respect to an increase of the pumping frequency (as long as the latter remains 
much slower than the relaxation time of the system $\Omega\ll\Gamma$),  the plateaux in the heat current are strongly affected by heat production in the leads. This is described by 
the second non-adiabatic corrections
to the heat currents $J_{\alpha}^{(2)}$, which account for the excess energy deposited into the leads as a consequence of the finite-time driving of the dots' levels.  

Heating effects are detrimental for the performance of  double-dot pump as a nanoscale engines. In fact, while the driven double dot can in principle be operated as  a nanoscale ``battery charger'' moving electrons from a lower to a higher chemical potential, or as a heat pump exchanging energy with two reservoirs with different temperatures, its efficiency is limited by dissipative effects due to leakage currents and finite-frequency operation. 
We show that despite these effects, the efficiency of a double-dot charge pump performing work against a dc-source can reach of up to 70\% of the  ideal value.

Our results are based on a generalized master equation approach in the regime of weak coupling  to the leads, which allows to take systematically into account the effects of a small but finite driving frequency. It is interesting to observe how fundamental thermodynamic relations, see e.g. Eq.\eqref{eq_work_heat}, naturally emerge from transport calculations  and how the adiabatic expansion allows to identify reversible and irreversible transport processes. 

In this work we focused in particular on the semi-classic regime of strong inter-dot coupling and large modulation amplitudes, which is the most interesting for applications of the double dot as a quantized charge source or, as we discussed above, as a nanoscale engine.  As future perspective, it will be interesting to extend our formalism to take into account coherences\cite{Leijnse08} between the double-dot states and cotunneling effects,\cite{Splett06} to address regimes where quantum effects play a major role for the dynamics of the system. 

\begin{acknowledgements}
 We thank J. Pekola, R. S\'anchez, and M. R. Wegewijs for fruitful discussions and we acknowledge funding by the Excellence Initiative of the German Federal and State Governments and by the Ministry of Innovation NRW. 
\end{acknowledgements}

\appendix

\section{Geometric properties of $I_{\alpha}^{(1)}$ and  $J_{\alpha}^{(1)}$ } \label{app_geometric}
It is well known that {\em adiabatic pumping} is of geometric nature, meaning that, in the limit of slow modulations $\Omega\to0$, the charge pumped through a system depends only on the specific shape of the path sustained by the system's parameters but not on its detailed time evolution.\cite{Avron00,Makhlin01}  The same holds true for the heat exchanged with the leads during one pumping cycle, as we discuss in this appendix. 

In terms of the expansion in Eq.~\eqref{eq:expansion}, the quantities that possess geometric properties are the terms to first order in the driving frequency $I_{\alpha}^{(1)}$ and $J_{\alpha}^{(1)}$, which are responsible for adiabatic pumping.  To emphasize this aspect, it is useful to express the charge and the heat that are exchanges with one lead during one pumping cycle, $Q^{(1)}_{\xi}=\int_{0}^{2\pi/\Omega}\! dt\, \xi^{(1)}(t)$ ($\xi\in\{ I_{\alpha}, J_{\alpha}\}$),  in terms of auxiliary vector fields in the space of the parameters.\cite{Calvo12} The key observation is that $I_{\alpha}^{(1)}$ and $J_{\alpha}^{(1)}$ are directly related to the time derivative of the instantaneous occupation probabilities, i.e. 
\begin{equation}\label{eq:A1}
I_{\alpha}^{(1)}(t)= {\bf e}\, \boldsymbol{\mathcal{I}}^{\alpha}_t \tilde{\W}_{t}^{-1}\frac{d}{dt}\P_{t}^{(0)}\equiv \boldsymbol{\varphi}^{I_{\alpha}}_{t}\frac{d}{dt}\P_{t}^{(0)},
\end{equation}
and similarly for the heat current $J_{\alpha}^{(1)}$. Here $\tilde{\W}_{t}^{-1}$ is a  pseudo-inverse of the evolution kernel ${\W}_{t}$.\cite{Calvo12} In the second identity, we introduced the vector-valued response coefficients $\boldsymbol{\varphi}^{\xi}_{t}$, which describes the rate at which charge ($\xi=I_{\alpha}$) or heat ($\xi=J_{\alpha}$) is transferred to lead $\alpha$ due to a change in the occupation probabilities.

The pumped ``charge'' $Q_{\xi}^{(1)}$ can then be written as a line-integral over a closed contour $C$ in the space of the driving parameters 
\begin{align}\label{eq:Qxi:A}
Q_{\xi}^{(1)}=\oint_{C}d\boldsymbol{\epsilon}\cdot\boldsymbol{\mathcal{A}}_{\xi}(\boldsymbol{\epsilon}), 
\end{align}
with   
\begin{equation}
\boldsymbol{\mathcal{A}}_{\xi}(\boldsymbol{\epsilon})=\sum_{i}\varphi^{\xi}_{i}(\boldsymbol{\epsilon})\boldsymbol{\nabla}p^{(0)}_{i}(\boldsymbol{\epsilon}).
\end{equation}
Here,  $\boldsymbol{\epsilon}=\sum_{\alpha}\epsilon_{\alpha}{\bf e}_{\alpha}$ indicates the ``position'' vector in the parameter space spanned by ${\bf e}_{\rm L}=(1,0)$ and ${\bf e}_{\rm R}=(0,1)$, and $\boldsymbol{\nabla}=\sum_{\alpha}{\bf e}_{\alpha}\partial_{\epsilon_{\alpha}}$. The vector field $\boldsymbol{\mathcal{A}}_{\xi}$ can be interpreted as a pseudovector potential defined in the space of the driving parameters, and its components are directly related to the concept of emissivity.\cite{Calvo12,Buttiker94}  The line integral on the right hand side of  Eq.\eqref{eq:Qxi:A} is independent of  how fast the orbit $C$ is traversed (provided that the adiabatic approximation holds), and therefore the pumped charge (or heat) does not depend on the driving frequency.

Extending the two-dimensional parameter space to a third dimension and using Stoke's theorem, Eq.\eqref{eq:Qxi:A} can be written as the surface integral 
\begin{align}\label{eq:Qxi:B}
Q_{\xi}^{(1)}= \iint_{\Sigma}d\boldsymbol{S}\cdot\boldsymbol{\mathcal{B}}_{\xi}(\boldsymbol{\epsilon}), 
\end{align}
where $\boldsymbol{\mathcal{B}}_{\xi}(\boldsymbol{\epsilon})=\boldsymbol{\nabla}\times \boldsymbol{\mathcal{A}}_{\xi}(\boldsymbol{\epsilon})$,  $\Sigma$ is the area  encircled by $C$ and $d\boldsymbol{S}$ the directed surface element in parameter space.    $Q_{\xi}^{(1)}$ can then be seen as the flux generated by the pseudo magnetic field $\boldsymbol{\mathcal{B}}_{\xi}$. The advantage of this representation is that  $\boldsymbol{\mathcal{B}}_{\xi}$ anticipates the conditions for finite pumping without referring to the specific details of the modulation.\cite{Calvo12}

We stress however that the geometric interpretation of the pumped charge (heat) holds only to first order in the driving frequency and cannot be generalized to higher-order contributions, which depend sensitively on the details of how the pumping orbit is traversed.

\section{Small-amplitude pumping in first and second order in $\Omega$}\label{app_secondorder}

\begin{figure}[ht]
\includegraphics[width=3.in]{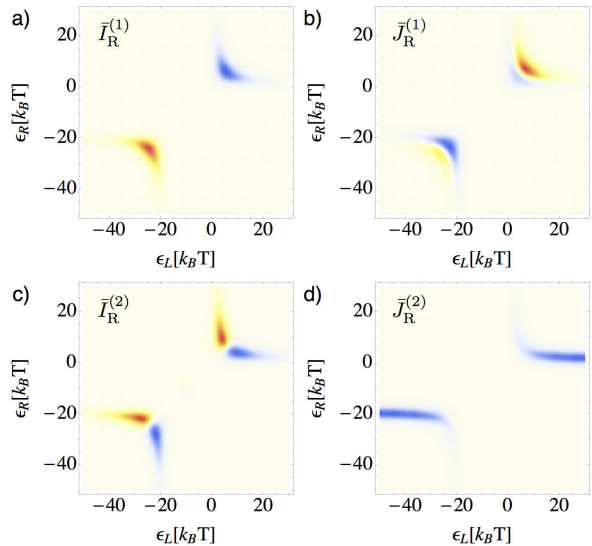}
\caption{(Color online) 
Density plots of pumped charge and heat currents to first and second order in the pumping frequency, as a function of the energy level of the left and right dot.
(a) Charge current to first order in $\Omega$, $\bar{I}_{\rm R}^{(1)}[e\Omega\frac{\pi\delta\epsilon_{\rm d}^2}{\left(k_\mathrm{B}T\right)^2}]$;  (b)  Heat current to first order in $\Omega$, $\bar{J}_{\rm R}^{(1)}[\Omega\frac{\pi\delta\epsilon_{\rm d}^2}{k_\mathrm{B}T}]$; (c)  Charge current to second order in $\Omega$,  $\bar{I}_{\rm R}^{(2)}[e\Omega\frac{\pi\delta\epsilon_{\rm d}^2}{\left(k_\mathrm{B}T\right)^2}]$,  and (d) heat current to second order in $\Omega$, $\bar{J}_{\rm R}^{(2)}[\Omega\frac{\pi\delta\epsilon_{\rm d}^2}{k_\mathrm{B}T}]$. In all panels:  $T_{\rm L}=T_{\rm R}=T$, $V=0$ and $U=20 k_{\rm B}T$, $t_{\rm c}=10 k_{\rm B}T$, $\Gamma_{\rm L}=\Gamma_{\rm R}=\Gamma/2$, $\hbar\Gamma=k_{\rm B}T/4$.
\label{fig_small_amp}}
\end{figure}
In order to get a better understanding of the characteristics of the pumped charge and heat, we present in this appendix some results in the regime of weak pumping. This means that the amplitudes of the parameter modulation are chosen to be small with respect to the extension of the triple points. We furthermore take the amplitudes of the time-dependent part of both energy levels $\epsilon_\mathrm{L}(t)=\bar{\epsilon}_\mathrm{L}+\delta \epsilon_\mathrm{L}\sin(\Omega t)$ and $\epsilon_\mathrm{R}=\bar{\epsilon}_\mathrm{R}+\delta\epsilon_\mathrm{R}\sin(\Omega t+\phi)$ to be equal, $\delta\epsilon_\mathrm{L}=\delta\epsilon_\mathrm{R}\equiv\delta\epsilon_\mathrm{d}$. Finally we are interested in the parameter cycles for which adiabatic charge and energy pumping is optimal and we therefore choose the phase difference between the parameters to be $\pi/2$. These conditions result in modulation cycles of circular shape in the stability diagram. The contributions to the pumped charge in the small-amplitude regime in first order in the frequency are proportional to the area of the enclosed cycle
\begin{eqnarray}\nonumber
\int_0^{2\pi/\Omega}\epsilon_\mathrm{L}(t)\dot{\epsilon}_\mathrm{R}(t)dt  = -\int_0^{2\pi/\Omega}\dot{\epsilon}_\mathrm{L}(t)\epsilon_\mathrm{R}(t)dt= \pi\delta\epsilon_\mathrm{d}^2.
\end{eqnarray}
 In order to further simplify our considerations we assume symmetric coupling, $\Gamma_\mathrm{L}=\Gamma_\mathrm{R}$ between the dots and their neighboring leads. Asymmetric couplings lead to qualitatively similar results as the ones presented here, except for slight changes in the symmetry of the pumped charge and heat as a function of the working point $(\bar{\epsilon}_\mathrm{L},\bar{\epsilon}_\mathrm{R})$.
 
The results for the pumped charge and energy current in first order in the frequency are shown in Fig.~\ref{fig_small_amp}a,b. Both of these currents are pure \textit{transport} properties, in the sense that we have $\bar{I}^{(1)}_\mathrm{L}=-\bar{I}^{(1)}_\mathrm{R}$ as well as $\bar{J}^{(1)}_\mathrm{L}=-\bar{J}^{(1)}_\mathrm{R}$.
Both charge and heat currents have contributions in the vicinity of the triple points in the stability diagram of the double dot. The pumped charge changes sign between the two triple points, which is due to the sign difference in the dependence  on the detuning of the effective coupling of the two hybrid states, $\Gamma_\mathrm{b/a}$, see Eq.~(\ref{eq:effective_coupling}).

The heat current has a different behavior: the contributions to the heat current  have a node in the central regions of the triple points. This is due to the fact that the transported energy becomes zero when the quantum dot level through which transport takes place is exactly at resonance. Furthermore we note that there is an asymmetry in the magnitude of the positive and the negative contribution. This is due to the fact that the maximum contribution to transport is shifted by a temperature-dependent factor, $k_\mathrm{B}T \ln 2$, with respect to the zero of the transported energy. 

Finally we note that in regime of small pumping-amplitudes, the average charge and heat currents to first order in $\Omega$ are directly proportional to the intensity of the pseudo-magnetic fields introduced in Appendix~\ref{app_geometric}, i.e. ${\bar{I}_{\alpha}^{(1)}\propto \mathcal{B}_{I_{\alpha}} }$ and ${\bar{J}_{\alpha}^{(1)}\propto \mathcal{B}_{J_{\alpha}} }$, so that the plots of Fig.~\ref{fig_small_amp}a,b represents a color-scale map of $\mathcal{B}_{I_{\rm R}}$ and $\mathcal{B}_{J_{\rm R}}$ for the case $V=0$ and $T_{\rm L}=T_{\rm R}$.    

We now consider the second order in $\Omega$ contribution to charge and heat transport. The pumped charge current $\bar{I}^{(2)}$ is -- due to charge conservation -- again a purely transported quantity with $\bar{I}^{(2)}_\mathrm{L}=-\bar{I}^{(2)}_\mathrm{R}$. In contrast to the current in first order in $\Omega$, this contribution is symmetric as a function of the mean energy $\bar{E}$ around $\bar{E}=-\frac{U}{2}$, and it is antisymmetric as a function of the detuning $\bar{\epsilon}$ around zero detuning. Another important difference is that it is due to \textit{single-parameter} pumping, indeed we find
\begin{eqnarray}
\bar{I}^{(2)} & = & e \frac{\Omega}{2\pi}\int_0^{2\pi/\Omega}\e(t)\ddot{\e}(t) dt \nonumber\\
&& \cdot \sum_{\eta=\mathrm{b,a}}\frac{d}{d\e}\left(\frac{\Gamma_{\mathrm{R}\eta}}{\Gamma_\eta}\right)\frac{d}{d\e}\left(p_\eta^{(0)}+p_\mathrm{d}^{(0)}\right)\bigg|_{\bar{\e}}.
\label{eq_charge_second_weak}
\end{eqnarray}
Charge pumping due to a single time-dependent parameter is not possible in the adiabatic regime. The different behavior of the second-order in frequency contribution can be understood by considering in detail a charge transfer process. Let us concentrate on the situation where the bonding level is close to resonance. The energy of the bonding level is \textit{decreased} whenever the absolute value of the detuning is \textit{increased} (the opposite effect is true for the antibonding state). A thereby induced \textit{loading} of the bonding state takes place either mostly through the left or mostly through the right lead depending on the effective coupling $\Gamma_{\mathrm{L/R},\mathrm{b}}$. The fast change of the detuning has the consequence that the effective coupling ``seen" by the tunneling particles is slightly delayed, leading to a slightly increased (decreased) coupling to the left lead with respect to the equilibrium situation when the detuning is increased (decreased), and vice versa for the coupling to the right lead  (the opposite effect is true for the antibonding state). This allows for the appearance of single-parameter pumping, as reflected in the current formula, Eq.~(\ref{eq_charge_second_weak}). However, since this delay in the effective couplings is equal for a detuning of equal magnitude and opposite sign, charge pumping takes place only if the average detuning $\bar{\epsilon}$ is different from zero, see Fig.~\ref{fig_small_amp}c.

Finally, we want to address the heat current in second order in the frequency, which is due to the finite-frequency operation. This heat current is \textit{generated} by the time-dependent fields and therefore has the same negative sign in both leads. In other words, due to the time-dependent modulation, heat is flowing from the double-dot system into the leads. In Fig.~\ref{fig_small_amp}d, the heat current into the right lead is shown. It shows large contributions whenever the level through which transport takes place (namely the bonding level at the transition $0\leftrightarrow 1$ and the antibonding level at the transition $1\leftrightarrow 2$) is in resonance with the right lead. Since the heat current in second order in the frequency is \textit{not} a transported quantity in this regime, contributions also far away from the triple points are visible, as long as one of the levels is close to resonance. The analogous situation, with an inverted behavior with respect to the detuning, is observed for the heat current into the left lead. 

This obviously different behavior of the heat current in second order in the frequency shows that here the second-order contribution is often larger than the first-order in frequency contribution. It is therefore of high relevance to consider  contributions in second order in the frequency when studying heating effects. We however want to emphasize that the rigorous expansion in the driving frequency remains well justified, since all higher contributions in frequencies can be shown to be suppressed in powers of $\Omega/\Gamma$.


%

\end{document}